\documentclass{aa}    
\usepackage{graphicx} 
 
\begin{document} 
 
\thesaurus{3(
	11.09.1 Centaurus A; 
	11.09.2; 
	11.09.4; 
	11.10.1; 
	13.19.1) 
             } 
\title{First Detection of Molecular Gas in the Shells of CenA
\thanks{Based on observations made with the Swedish-ESO Submillimeter
Telescope (SEST) at La Silla, Chile}} 

\author{V. Charmandaris 
	\inst{1,2},  	
	F. Combes
	\inst{1},
	\and 
	J.M. van der Hulst
	\inst{3}
	} 
\offprints{V. Charmandaris, vassilis@astro.cornell.edu} 
  
\institute{
	Observatoire de Paris, DEMIRM, 61 Avenue de l'Observatoire, 
	F-75014 Paris, France 
	\and
	Cornell University, Astronomy Department, Ithaca, NY 14853-6801, USA
	\and
	Kapteyn Institute, University of Groningen, Postbus 800, 9700 AV, 
	Groningen, The Netherlands
	} 
\date{Received 13 January 2000 / Accepted 17 February 2000} 
\authorrunning{Charmandaris et al.} 
\titlerunning{First Detection of Molecular Gas in the Shells of CenA}
\maketitle 
 
\begin{abstract}   

Shells are faint arc-like stellar structures, which have been observed
around early type galaxies and are thought to be the result of an
interaction.  HI gas has recently been detected in shells, a
surprising result in view of the theoretical predictions that most of
the gas should decouple from stars and fall into the nucleus in such
interactions.  Here we report the first detection of molecular gas
(CO) in shells, found 15\,kpc away from the center of NGC\,5128 (CenA),
a giant elliptical galaxy that harbors an active nucleus (AGN). The
ratio between CO and HI emission in the shells is the same as that
found in the central regions, which is unexpected given the
metallicity gradient usually observed in galaxies.  We propose that
the dynamics of the gas can be understood within the standard picture
of shell formation if one takes into account that the interstellar
medium is clumpy and hence not highly dissipative. The observed metal
enrichment could be due to star formation induced by the AGN jet in
the shells. Furthermore our observations provide evidence that
molecular gas in mergers may be spread out far from the nuclear
regions.
\end{abstract} 
  
\section{Introduction} 

Early type galaxies (E/S0) are often found to be surrounded by faint
arc-like stellar structures, called shells or ripples (\cite{mc}), due
to the accretion and subsequent merging of a smaller companion galaxy.
It is widely accepted that the high frequency of galaxies with shells
($\sim$\,50\%) attests to the importance of merging in galaxy
formation (Schweizer \& Seitzer 1988, 1992). Simulations of the {\em stellar
component} have shown that the shells or ripples are created either by
``phase-wrapping'' of the tidal debris of the accreted companion on
nearly radial orbits (\cite{quinn}), or by ``spatial-wrapping'' of
matter in thin disks (\cite{dupraz}, \cite{hq}). 


CenA is a giant elliptical galaxy with strong radio lobes on either
side of a prominent dust lane situated along its minor axis
(\cite{clarke}). Additionally, optical and HI observations
(\cite{dufour}, \cite{vangorkom}) show a warped gaseous disk which has
been accreted along the minor axis of this apparently prolate
elliptical galaxy. CO mapping suggests that the disk contains 2$\times
10^8$ M$_{\odot}$ of molecular gas (\cite{eckart}).  Recently, mid-IR
observations revealed the presence of a bisymmetric bar-like
distribution of hot dust in the inner disk (\cite{mirabel}). High
contrast optical images of the galaxy show stars distributed in a
large number of faint narrow shells around the galaxy (\cite{malin}).
The presence of the warped gas disk and shells suggests that CenA has
accreted one (or more) smaller disk galaxy(ies) approximately $\sim
10^8$ yrs ago (\cite{quillen}).

Schiminovich et al. 1994 detected 4$\times 10^8$ M$_{\odot}$ of HI gas
associated with the stellar shells, having the same arc-like curvature
but displaced 1 arcmin to the outside of the stellar shells. This
result is intriguing since in general it is thought that the dynamics
of the gas and stellar components are decoupled during a merging
event. Detailed numerical modeling of the infall of a small companion
on a massive elliptical has demonstrated that aligned and interleaved
shells are formed through phase wrapping of the companion's stars on
almost radial orbits (\cite{quinn}, \cite{dupraz}). When gas is taken
into account, due to its dissipation it rapidly concentrates in the
nucleus and does not form any shell (\cite{weil}).  Another
possibility is that shells result from space wrapping when the
relative angular momentum of the two galaxies is high.  In this case
the stellar shells do not have the same regular structure
(\cite{prieur}) and as they rotate around the central potential they
dissolve more rapidly. The gas could remain associated with the
stellar shells during a few dynamical times before condensing to the
center.  The morphology of the shells in CenA, though, suggests a
combination of both phase and space wrapping since there are both a
number of shells aligned with the major axis of the prolate giant
elliptical and there are a few which are irregular. Furthermore, the HI
is mostly associated with what Malin et al. (1983) called the diffuse
shells.

To explain the presence of gas in phase-wrapped shells, one should
consider the interstellar medium as multiphase: a large fraction of
the ISM could be composed of dense clumpy material with low
dissipation.  During a galaxy merger the dense gas behaves almost as
collisionless particles and can orbit through the center as the stars
do. To trace this dense component we attempted to detect molecular gas
from the shells in CenA.  The results were positive beyond our
expectation, as described now.

\begin{figure*}
\resizebox{\hsize}{!}{\includegraphics{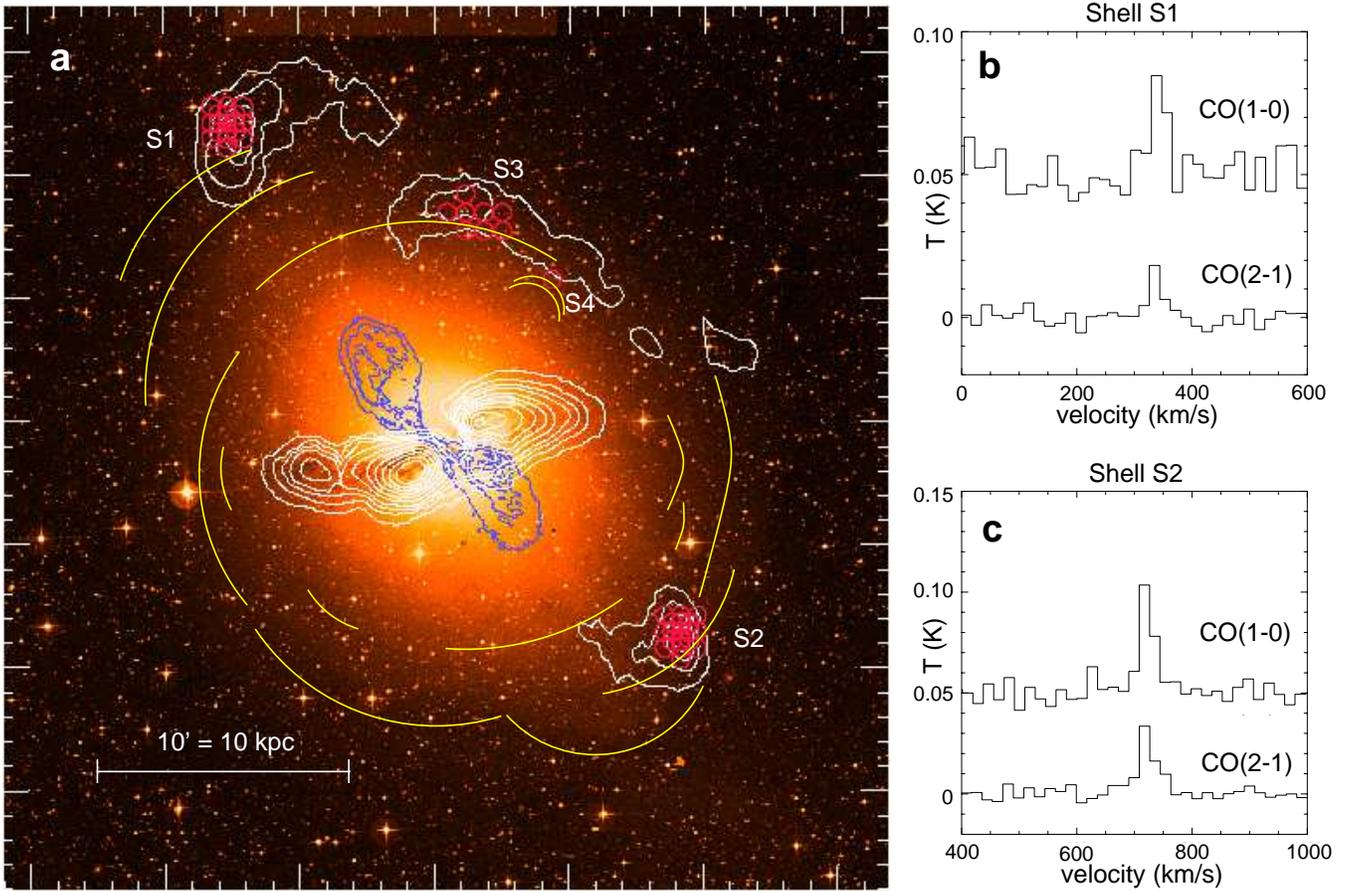}}  
 \caption{a) A Digitized Sky Survey optical image of CenA with the
 contours of HI gas (from \cite{david1}) superimposed in white. The HI
 contour levels are 1, 4, 7, 10, 15, 30, 35, 40\,$\times$\,10$^{20}$ cm
 $^{-2}$. North is up and east is to the left, while the image scale is
 shown by the horizontal bar. The positions observed in CO are marked
 with the red circles whose size corresponds to the SEST 44$''$ beam
 of CO(1-0). The type of each map (half-beam spacing or simple
 pointing) is evident by the placement of the circles. The locations of
 the outer stellar shells are underlined by the yellow solid lines
 (see also Fig. 1a of Schiminovich et al 1994) .  The inner 6cm radio
 continuum lobes (from \cite{clarke}) are depicted by the blue contours
 (contour levels 0.01, 0.05, 0.1 Jy/beam). Note the jet alignment with the
 location of the CO detections. The outer radio lobes are far more
 extended.  b) CO(1-0) and CO(2-1) spectra towards the northern shell
 S1 with the temperature scale in main beam T$_{\rm mb}$, smoothed to
 18\,km\,s$^{-1}$.  c) Same as in b) but for the southern shell S2.  }
 \label{fig}
\end{figure*}


\section{Observations and Data Reduction}   

The observations have been carried out in May 1999 in La Silla, Chile,
with the 15m Swedish-ESO Submillimeter Telescope (SEST)
(\cite{booth}).  We used the IRAM 115 and 230 GHz receivers to observe
simultaneously at the frequencies of the $^{12}$CO(1--0) and the
$^{12}$CO(2--1) lines.  At 115 GHz and 230 GHz, the telescope
half-power beam widths are 44$''$ and 22$''$, respectively.  The
main-beam efficiency of SEST is $\eta$$_{\rm mb}=T_{\rm A}^*/T_{\rm
mb}$=0.68 at 115 GHz and 0.46 at 230 GHz (SEST handbook, ESO).  The
typical system temperature varied between 300 and 450 K (in $T_{\rm
A}^*$ unit) at both frequencies. A balanced on-off dual beam switching
mode was used, with a frequency of 6 Hz and two symmetric reference
positions offset by 12$'$ in azimuth.  The pointing was regularly
checked on the SiO maser R\,Dor as well as using the continuum emission
of the nucleus of CenA.  The pointing accuracy was 4$''$ rms.  The
backends were low-resolution acousto-optical spectrometers.  The total
bandwidth available was 500 MHz at 115 GHz and 1 GHz at 230 GHz, with
a velocity resolution of 1.8 km s$^{-1}$.  We mapped four regions of
CenA associated with HI and stellar shells (noted as S1--4 in
Fig.~\ref{fig}a).  Regions S1 and S2 were covered with a 3$\times$3
half CO(1-0) beam maps centered at $\alpha$=13h26m16.1s,
$\delta$=$-42^{\circ}$46$'$55.7$''$ and $\alpha$=13h24m35.4s,
$\delta$=$-42^{\circ}$08$'$34.9$''$ (J2000) respectively. S3 consisted
of a series of pointings along an optical shell and S4 was a single
pointing. The rms noise per pointing was $\sigma_{\rm mb}\sim$\,3\,mK for
both frequencies.

\section{Results and Discussion}

We detected CO emission from two of the fully mapped optical shells
(S1 and S2) with associated HI emission, indicating the presence of
4.3$\times 10^7$ M$_{\odot}$ of H$_2$ assuming the standard CO to
H$_2$ conversion ratio. The CO lines were detected at the 4$\sigma$
level in four out of the nine pointings of each 3$\times$3 map.
Figure~\ref{fig} shows an optical image of CenA, with the positions
mapped in CO, the location of the HI and stellar shells, as well as
the spectra of the strongest CO detections found at the central
position of each map.  The molecular gas in both positions is clearly
associated with the HI shells since their velocities
(340\,km\,s$^{-1}$ for S1 and 720\,km\,s$^{-1}$ for S2) follow the
velocities of the HI as presented in Fig.~3 of \cite{david1}. The width
of CO lines is $\sim$ 20 km\,s$^{-1}$ while in a beam of twice the
size the HI linewidth is 80 km\,s$^{-1}$. This difference is easily
explained by the systematic velocity gradients within the beams.

As shown in Table~\ref{tab}, the ratio between the H$_2$ mass (derived
from the CO emission) and the HI mass found in the same area is nearly
unity for both detected shells and central regions. In fact, the mass
ratio between H$_2$ and HI that we find in CenA is about normal for
giant spiral galaxies, where the global M(H$_2$)/M(HI) has been found
on average to be equal to unity in a survey of 300 objects
(\cite{young}).  According to the type of the galaxy and its star
formation activity as measured by the far-infrared flux, this ratio
could vary (\cite{sage}).  It has been found equal to
M(H$_2$)/M(HI)=0.2 in the Coma supercluster (\cite{casoli}), and it is
even much lower (by a factor 10) in dwarf galaxies (\cite{taylor}).

\vspace*{-0.5cm}
\begin{table}[!h]
\caption[]{HI and CO gas properties}
\begin{tabular}{cccc}  
\hline 
                        & Center
                        & Shell S1 
                        & Shell S2 \\
M(HI)$^{\mathrm{a}}$ (M$_{\odot}$)         & 3.5$\times10^{8}$
                        & 2.14$\times10^{7}$
                        & 2.17$\times10^{7}$\\
M(H$_{2}$) (M$_{\odot}$)& 3.3$\times10^{8}$
                        & 1.7$\pm 0.1\times10^{7}$
                        & 2.2$\pm 0.1\times10^{7}$\\
M(HI)/M(H$_{2}$)        & 1.06 & 1.25   & 0.99 \\
CO(2-1)/CO(1-0) &  0.5$^{\mathrm{b}}$           & 0.55  & 0.75 \\
\hline
\end{tabular} 
\label{tab} 
\begin{list}{}{}
\item[$^{\mathrm{a}}$] Based on the maps of \cite{david1}.
\item[$^{\mathrm{b}}$] Using the published values of \cite{eckart}.
\end{list}
\vspace*{-0.5cm}
\end{table}

The total gas mass found in CenA is almost 10$^9$ M$_\odot$,
comparable to that of a giant spiral galaxy. Since this gas must have
once belonged entirely to the accreted companion, we can deduce that
the latter was not a dwarf, but a massive spiral such as the Milky
Way. This also explains the large ratio of CO emission to HI gas
observed, leading through the standard conversion ratio to the high
values of M(H$_2$)/M(HI) found. More surprising is the fact that this
ratio is the same (close to unity) in the center, and at 15\,kpc from
it. In general for most galaxies, due to the metallicity gradient
(\cite{vila}), the ratio decreases exponentially with radius from
about 30 in the central part to less than 0.1 in the outermost parts
where CO is detected (\cite{combes}).  Furthermore the CO(2-1)/CO(1-0)
ratio in the shells, corrected for the different beam sizes, is $\sim$0.6.
This is not much lower than the 0.9 observed in the nuclei of nearby spirals
(\cite{braine}) and slightly above what is found in disks. This means
that CO lines are not highly subthermally excited and therefore the
density of the gas is at least 10$^{4}$\,cm$^{-3}$.

Our data reveal the presence of dense molecular gas in the
shells. This presence helps to understand the existence of HI shells,
the HI gas being the diffuse envelopes of the dense molecular clumps,
the interface between the interstellar radiation field and the clumps.
Should this diffuse gas be present alone, it would have been driven
quickly towards the center during the stellar shells formation.
Modeling of the ISM should take into account a multiphase medium with
a low dissipation gaseous component.  Such models have been developed
by numerically simulating the gas dynamics through a cloud-collision
scheme (\cite{kojima}, \cite{cc}). In these simulations the low
dissipation of gas enables a fraction of it to follow the stellar
component.  However, due to a different initial distribution of gas
and stars in the companion galaxy, the gaseous shells do not coincide
with the stellar ones. More precisely the gas is radially more
extended in the companion disk than the stars and therefore less
gravitationally bound. Hence, during the merger the gas is the first
to be tidally stripped from the companion and thus does not experience
dynamical friction. On the other hand, dynamical friction brakes
efficiently the remaining stellar core which is tidally stripped
somewhat later in the merging process. Since the stars have lost more
energy than the gas they oscillate in the potential with smaller
apocenters and thus the shells they form are located inside the gas
shells. In this framework the observed displacement of the gas with
respect to stars in CenA shells is naturally explained. The presence
of a diffuse gas (seen in HI) at those regions is expected since part
of the clumpy gaseous component dragged into shells will be
dissociated either by local star formation activity or by the global
galactic radiation field.

What remains unclear is {\em why the metallicity in the shells is
sufficient for the CO emission to be detected}. Indeed as described
above, the dynamical friction segregates the gas in the elliptical
potential according to its initial distribution in the companion and a
metallicity gradient should still exist in the final merger remnant.

A solution to this puzzle could be found if we consider the effect of
the radio jet of CenA on the gaseous shell. Note that from the four
shells mapped in CO emission only the two shells (S1, S2 in
Fig.~\ref{fig}) {\em aligned with the jet} have been detected.
Optical filaments are observed along the jet, only in the regions
where the jet and the shells intersect, suggesting that the ambient
gas is ionized by the nuclear beamed radiation or is excited by shocks
(\cite{graham81}, \cite{morganti}). Moreover, a group of blue stars
has been discovered, just between the northern HI shell and its
corresponding outermost optical filament (see Fig.~1 and 2 in Graham
1998 and Fasset \& Graham 2000). The formation of these stars is proposed to be
triggered by the impact of the radio jet on the HI shell. The HII
regions ionized by these blue stars have measured velocities
coinciding with those of the adjacent HI and CO gas. A study of the
stellar content reveals several generations of stars whose lifetimes
extend over a period of 15$\times 10^{6}$ years, while supernovae
dissipate into the surrounding medium in less than 10$^{6}$ years
(\cite{graham99}). This observed stellar activity has certainly
enriched the observed gas in metals and can explain our detection of
CO molecules.  Moreover, the impact of the radio jet could be
responsible for the formation of new dense molecular clouds. It is
necessary, however, that some of the gas was already present in the
jet region, and this gas be driven there by the shells.  The formation
of gaseous shells requires the presence of dense molecular gas first,
since the jet can only maintain or form secondary molecular clouds (as
quoted by \cite{graham98}), but cannot serve as their primary
formation mechanism. Indeed, we can eliminate the alternative
possibility in which diffuse gas is spread by the interaction
everywhere, and is compressed in molecular clouds in the jet; there
would then be no coincidence between the HI gas and the stellar shells
in this scenario.

\section{Conclusions} 

We have detected CO molecules in two shells aligned along the major
axis of Centaurus A. The molecular gas is globally associated with the
HI and stellar shells but with a radial shift, the HI being the more
external component, and the stars the more internal.  The presence of
molecular clouds in these distant shells is compatible with the
dynamical scenario of phase-wrapping, following the merger of a spiral
galaxy with Centaurus A.  Part of the interstellar medium of the
spiral is clumpy, with very low collision rate and dissipation, and
can follow nearly radial orbits during the merger, like the stellar
component, without accumulating towards the center. The differential
dynamical friction experienced by the gas and stellar components, that
are unbound from the spiral companion at different epochs, can explain
the radial shifts between the different shells.
  
The detection of CO emission far from the center implies the presence
of H$_2$ molecules as far as 1.16\,R$_{25}$.  The present detections,
taking into account that only a small fraction of the shells was
mapped in CO, suggest that more than 50\% of the gas in the outer
regions of CenA is in molecular form, and at least 10\% of the total
molecular gas detected in CenA is not in the nucleus. Moreover the
derived HI/H$_2$ mass ratio is nearly constant with radius. This
requires a metallicity enrichment in the most external gas, that could
be due to the interaction between the gaseous shells and the radio
jet. This prototypical example of a gaseous accretion suggests that
the molecular gas is not always confined in the nuclear regions in
merger remnants.

\begin{acknowledgements}
The authors are grateful to J. van Gorkom for providing the HI data of
Cen~A, as well F. Mirabel, M. Noguchi, K. Uchida and an anonymous referee
for their valuable comments.  VC would like to acknowledge the financial support
from a Marie Curie fellowship (TMR grant ERBFMBICT960967).
\end{acknowledgements} 


\end{document}